\documentclass[prd,superscriptaddress,showpacs,nofootinbib,amsmath,amssymb,aps,10pt]{revtex4}

\usepackage{bm}
\usepackage{amsfonts}
\usepackage{latexsym}
\usepackage[latin1]{inputenc}
\usepackage{graphicx}
\usepackage{amsmath}
\usepackage{palatino}
\usepackage{mathpazo}
\usepackage{epstopdf}
\usepackage{booktabs}
\usepackage{dcolumn}
\usepackage{array}

\usepackage{wrapfig}
\usepackage{wasysym}

\def\jnl@style{\it}
\def\aaref@jnl#1{{\jnl@style#1}}

\def\aaref@jnl#1{{\jnl@style#1}}

\def\aj{\aaref@jnl{AJ}}                   
\def\apj{\aaref@jnl{ApJ}}                 
\def\apjl{\aaref@jnl{ApJ}}                
\def\apjs{\aaref@jnl{ApJS}}               
\def\apss{\aaref@jnl{Ap\&SS}}             
\def\aap{\aaref@jnl{A\&A}}                
\def\aapr{\aaref@jnl{A\&A~Rev.}}          
\def\aaps{\aaref@jnl{A\&AS}}              
\def\mnras{\aaref@jnl{Mon.~Not.~Roy.~Astron.~Soc.}}             
\def\prd{\aaref@jnl{Phys.~Rev.~D}}        
\def\prc{\aaref@jnl{Phys.~Rev.~C}}  
\def\prl{\aaref@jnl{Phys.~Rev.~Lett.}}    
\def\qjras{\aaref@jnl{QJRAS}}             
\def\skytel{\aaref@jnl{S\&T}}             
\def\ssr{\aaref@jnl{Space~Sci.~Rev.}}     
\def\zap{\aaref@jnl{ZAp}}                 
\def\nat{\aaref@jnl{Nature}}              
\def\aplett{\aaref@jnl{Astrophys.~Lett.}} 
\def\apspr{\aaref@jnl{Astrophys.~Space~Phys.~Res.}} 
\def\physrep{\aaref@jnl{Phys.~Rep.}}      
\def\physscr{\aaref@jnl{Phys.~Scr}}       
\def\commat{\aaref@jnl{Comm.~Math.~Phys.}}              
\def\science{\aaref@jnl{Science}}               
\def\cqg{\aaref@jnl{Classical Quant.~Grav.}}            
\def\jpcs{\aaref@jnl{JPCS}}                                     
\def\ijmpd{\aaref@jnl{Int.~J.~Mod.~Phys.~D}}                    
\def\grg{\aaref@jnl{Gen.~Relat.~Gravit.}}               
\def\rpp{\aaref@jnl{Rep.~Prog.~Phys.}}          
\def\npa{\aaref@jnl{Nucl.~Phys.~A}}        
\def\lrr{\aaref@jnl{Living Rev.~Rel.}}                   
\def\jcap{\aaref@jnl{J.~Cosmology Astropart.~Phys.}}    
\def\rmp{\aaref@jnl{Rev.~Mod.~Phys.}}   

\allowdisplaybreaks[1]

\begin{document}

\title{Accretion disks around neutron and strange  stars in $\mathcal{R}^2$ gravity}

\author{Kalin V. Staykov}
\email{kstaykov@phys.uni-sofia.bg}
\affiliation{Department of Theoretical Physics, Faculty of Physics, Sofia University, Sofia 1164, Bulgaria}

\author{Daniela D. Doneva}
\email{daniela.doneva@uni-tuebingen.de}
\affiliation{Theoretical Astrophysics, Eberhard Karls University of T\"ubingen, T\"ubingen 72076, Germany}
\affiliation{INRNE - Bulgarian Academy of Sciences, 1784  Sofia, Bulgaria}

\author{Stoytcho S. Yazadjiev}
\email{yazad@phys.uni-sofia.bg}
\affiliation{Department of Theoretical Physics, Faculty of Physics, Sofia University, Sofia 1164, Bulgaria}


\begin{abstract}

We study the electromagnetic spectrum of accretion disks around neutron and strange stars in $\mathcal{R}^2$ gravity. Both static and rapidly rotating models are investigated. The results are compared with the General Relativistic results. We found difference between the results in both theories of about 50\%  for the electromagnetic flux and about 20\% in the luminosity for models with equal mass and angular velocity in both theories. The observed differences are much lower for models rotating with Kelperian velocity and with equal masses.

\end{abstract}
\maketitle
\date{}
\section{Introduction}

In weak field regime General relativity (GR) is extensively  tested and confirmed, but this is not the case for the strong field regime. There are theoretical and observational concerns about viability of the theory for strong gravitational fields (around black holes and neutrons stars) and on cosmological scales. A major problem is the recently discovered accelerated expansion of the Universe. To be explained in the context of GR an exotic matter, dark energy, constituting 73 \% of the mass-energy content of the Universe, should be introduced. However, there is an alternative way based on the idea that we do not fully understand gravity, therefore a modification to the Einstein equations should be made. A class of such alternative gravitational theories are the so called $f(\mathcal{R})$ theories in which the General Relativistic Lagrangian, the Ricci scalar $\mathcal{R}$, in the Einstein-Hilbert action is replaced with a more general one -- a function of the Ricci scalar $f(\mathcal{R})$ \cite{Sotiriou2010,DeFelice2010a}.

Determining the parameters of a compact star is made mainly by optical observations which are quite inaccurate because of the size of those objects. This, combined with the uncertainty in the equation of state (EOS), leads to a major problem in testing alternative theories of gravity. Therefore, we should search for additional manifestations of the specific alternative theory of gravity we are interested in testing.  A good tool for this purpose could be  the electromagnetic spectrum of an accretion disk around compact object.

The first comprehensive study of accretion disks,  using a Newtonian approach, was made in  \cite{Shakura1973}. Latter a General Relativistic model of thin accretion disk was developed in \cite{Novikov1973}. The model was further studied in \cite{Page1974} and \cite{Thorne1974}. The thin accretion disk model is relatively simple. It suggest that the particles  are moving on geodesics (on Keplerian orbits), but for this to be true the central object should have weak magnetic field, otherwise the orbits in the inner edge of the disk will be deformed.  The disk is supposed to be in  steady-state -- the mass accretion rate is constant in time and does not depend of the radius of the disk. The disk is supposed to be in hydrodynamical and thermodynamic equilibrium which imply that it has a black body electromagnetic spectrum and properties. 

General Relativistic temperature profiles and spectra  from a thin accretion disk around neutron stars are calculated in \cite{Bhattacharyya2000, Bhattacharyya2001a, Bhattacharyya2001b, Bhattacharyya2002}, and temperature profiles of accretion disks around strange stars are calculated in \cite{Bhattacharyya2001}. The electromagnetic properties of the accretion disk around  neutron, strange and boson stars and comments on the possibility of determining the nature of the central object from the disk's spectrum are made in \cite{Kovacs2009} and \cite{Duanilua2015}.  The emissivity properties of a thin accretion disk have been  investigated for more exotic compact objects too. Rotating and non-rotating boson and fermion stars, gravastars and wormholes were investigated in \cite{Torres2002,Yuan2004,Guzman2004,Kovacs2009a,Bambi2013,Bambi2013a,Kong2014}. Some alternative models like brane world modes, Horava-Lifshitz gravity and Chern-Simons gravity, have been investigated as well \cite{Pun2008a,Harko2009, Harko2010,Harko2011}. In \cite{Pun2008} the authors study thin accretion disks in vacuum $f(\mathcal{R})$ gravity in static and spherically symmetric spacetime, i.e. accretion disk around  static and spherically symmetric black holes. They have found that a particular signature in the electromagnetic spectrum of the disks can appear and according to their conclusions that will allow the alternative theory to be tested. A relation between disk electromagnetic properties and the multiple moments of the central object are investigated in \cite{Pappas2015b}. In this paper the author discuses some independent of the equation of state relations for the electromagnetic spectrum of the disk. In most of the above mentioned papers different effects are neglected, like the light bending etc. A detailed calculation and analysis including all these effects one can find in \cite{Li2005}.

In the current paper we will be interested in accretion disks around neutron and strange stars in $\mathcal{R}^2$ gravity. We will investigate both static and rapidly rotating models and the electromagnetic properties of the disk around those compact objects. The scope of our study will be restricted to the differences that will occur compared with the GR case, so we will restrict ourself to small number of favoured by the observations EOS.

The structure of this paper is as follow. In Section II we present the basic steps for deriving the angular velocity, the specific energy and the specific angular momentum of a particle moving on a circular orbit around compact object. Deriving the innermost stable circular orbit (ISCO) is also discussed. In Section III we comment on the structure and the electromagnetic properties  of a thin accretion disk around compact object. The basic equations are presented and discussed. In Section IV we present our results. It starts with general discussion and contains three subsections concerning the  static case,  models rotating with equal angular velocities and models rotating with Keplerian velocity. The paper ends with conclusions.

\section{Particle on a circular orbit around compact star}

In this section we briefly present the basic mathematical equations for determining the angular velocity $\Omega_p$, the specific energy per unit mass $E$, and the specific angular $L$ momentum per unit mass along the axis of symmetry of a particle on a circular orbit in the gravitational potential of a compact object. The derivation of the innermost stable circular orbit (ISCO) is considered too.

We are considering  a stationary and axisymmetric spacetime with metric:

\begin{eqnarray} \label{Metric}
ds^2 = g _{tt}dt^2 + g_{rr}dr^2 + g_{\theta \theta}d\theta^2 + 2g_{t\varphi}dtd\varphi + g_{\varphi\varphi}d\varphi^2.
\end{eqnarray}
All the metric functions depend only on the coordinates $r$ and $\theta$. Because of the stationary and axial Killing symmetries of the metric, there are  two constants of motion, namely $E=-u_t$ and $L=u_{\varphi}$, where $u^{\mu} = \dot{x}^{\mu} = dx^{\mu}/d\tau$ is the four-velocity of the massive particle moving in the gravitational field of the central object. The two conservation laws can be written  in the following way

\begin{eqnarray}
\frac{dt}{d\tau} = \frac{Eg_{\varphi\varphi} + Lg_{t\varphi}}{g_{t\varphi}^2 - g_{tt}g_{\varphi\varphi}}, \\
\frac{d\varphi}{d\tau} = -\frac{Eg_{t\varphi} + Lg_{tt}}{g_{t\varphi}^2 - g_{tt}g_{\varphi\varphi}}.
\end{eqnarray}
Here $t$  denotes the coordinate time, and  $\tau$  the proper time. From  the normalization condition for massive particle $g^{\mu\nu} u_{\mu}u_{\nu} = -1$, we  have

\begin{eqnarray} \label{eq:4v_norm}
g_{rr}\dot{r}^2 + g_{\theta\theta}\dot{\theta}^2  = V(r).
\end{eqnarray}

We are investigating a thin accretion disk in the equatorial plane, therefore  $\theta=\frac{\pi}{2}$, and  the problem reduces to an effective one dimensional problem
\begin{eqnarray}
g_{rr}\dot{r}^2 = V(r),
\end{eqnarray}
with an effective potential

\begin{eqnarray}
 V(r)=\left[-1 -  \frac{E^2g_{\varphi\varphi} + 2EL g_{t\varphi} + L^2 g_{tt}}{g_{t\varphi}^2 - g_{tt}g_{\varphi\varphi}}\right]. 
\end{eqnarray}

To determine the inner edge of the disk we need to determine the radius of the ISCO. A stable circular orbit with a radius $\bar{r}$ for a particle moving in the effective potential $V(r)$ is determined by the conditions $V(\bar{r})= 0 = V^{'}(\bar{r})$ and $V^{\prime\prime}(\bar{r})>0$, where the derivatives are with respect to the radial coordinate. The  condition $V^{\prime\prime}(\bar{r})=0$ gives the ISCO radius. The angular velocity $\Omega_p$ of a particle moving on a circular equatorial orbit is defined as $\Omega_{p}=\frac{u^\varphi}{u^{t}}=\frac{d\varphi}{dt}$ and it
can be found from  the geodesic equations. Finally we obtain the angular velocity the specific energy and the specific angular momentum:

\begin{eqnarray}
&\Omega_p = \frac{d\varphi}{dt} = \frac{-\partial_{r} g_{t\varphi} \pm \sqrt{(\partial_{r}g_{t\varphi})^2 - \partial_{r}g_{tt}\partial_{r}g_{\varphi\varphi}}}{\partial_{r}g_{\varphi\varphi}},\\
&E = \frac{-g_{tt} - g_{t\varphi}\Omega}{\sqrt{-g_{tt} - 2\Omega g_{t\varphi} 
- \Omega^2 g_{\varphi\varphi}}},\\
&L = \frac{g_{t\varphi} + g_{\varphi\varphi}\Omega}{\sqrt{-g_{tt} - 2\Omega g_{t\varphi} 
- \Omega^2 g_{\varphi\varphi}}}.
 \end{eqnarray}
 
The derived quantities depend only on the metric coefficients and we assumed a general form of a stationary axixymmetric  metric (\ref{Metric}). Therefore, the presented in this section equations are valid in GR as well as in $\mathcal{R}^2$ gravity.  
 
 \section{Thin accretion disk structure and electromagnetic properties}
 
 The relativistic thin accretion disk model is developed by the authors of  ~\cite{Novikov1973,Page1974, Thorne1974}. A sufficient condition for the disk to be thin is the radius of the disk to be much bigger than it's maximal half thickness $H$. In the thin disk model  it is supposed that the disk is in hydrodynamical and thermodynamic equilibrium and that it is radiatively efficient. The first condition allows us to describe the disk's electromagnetic properties as it is a black body, and the second one means that all the gravitational energy converted into heat by tensions is radiated from the disk and the disk's temperature does not increase. The disk is supposed to be in steady state -- the mass accretion rate is constant in time and does not depend of  the radius. 
 
 In the thin accretion disk model the particles are assumed to be moving at Keplerian orbits with angular velocity $\Omega_p $, specific energy $E$, and specific  angular momentum $L$, and they are slowly inspiraling to the central object. Because the disk is assumed to be thin we can use near equatorial plane approximation, and therefore  it is convenient to rewrite the metric (\ref{Metric}) in the following form:
 
 \begin{eqnarray} \label{Metric_equat}
 ds^2 = g _{tt}dt^2 + g_{rr}dr^2  + 2g_{t\varphi}dtd\varphi + g_{\varphi\varphi}d\varphi^2 + dz^2.
 \end{eqnarray}
 
The accreting disk matter  is described by the following energy-momentum tensor:
 
 \begin{equation}
 T^{\mu\nu} = \rho_0u^{\mu}u^{\nu} + 2u^{(\mu}q^{\nu)} + t^{\mu\nu}.
 \end{equation} 
Where $u^{\mu}$ is the four-velocity, $q^{\mu}$ is the energy flow four-vector, $\rho_0$ is the rest mass density, and $t^{\mu\nu}$ is the stress tensor. 

The structure equations for the disk can be derived from the energy, the angular momentum and the mass conservation laws.
The first structure equation comes from the mass conservation and it has the form 

\begin{equation}
\dot{M}_0 \equiv -2\pi\sqrt{-g}\Sigma u^{r} = const,
\end{equation}
where $u^{r}$ is the radial component of the four-velocity, $g$ is the determinant of the metric (\ref{Metric_equat}) in the equatorial plane, and $\Sigma$ is the surface density, defined as 

\begin{equation}
\Sigma = \int_{-H}^{H}\left<\rho_0\right>dz,
\end{equation}
and $\left<\rho_0\right>$  is the rest mast density averaged over time and azimuthal angle $\varphi$.

The second structural equation comes from the angular momentum conservation: 

\begin{equation}
\partial_{r}\left[\dot{M}_0L - 2\pi\sqrt{-g}W^{r}_{\varphi}\right] = 4\pi\sqrt{-g}FL,
\end{equation}
where $F$ is the radiated energy flux from one site of the disk, and $W^{r}_{\varphi}$ is the time-averaged torque per unit circumference at a particular radius and it is defined as 

\begin{equation}
W^{r}_{\varphi}(r) \equiv \int_{-H}^{H}\left<t^{r}_{\varphi}\right>dz.
\end{equation}

The last structure equation is derived from the energy conservation law and it is

\begin{equation}
\partial_{r}\left[\dot{M}_0E - 2\pi\sqrt{-g}\Omega W^{r}_{\varphi}\right] = 4\pi\sqrt{-g}FE.
\end{equation}

To derive the equations for the electromagnetic properties of the disk we will need the fundamental relation for circular geodesics $\partial_{r}E = \Omega \partial_{r}L$.

As we mentioned we assume that the disk is in termodinamical equilibrium therefore we can consider it to radiate as a black body. Then the electromagnetic flux from the disk can be derived from the structure equations and has the form: 

\begin{equation}
F(r) = -\frac{\dot{M}_0}{4\pi\sqrt{-g}}\frac{\partial_{r}\Omega}{\left(E- L\Omega\right)^2}\int_{r_{in}}^{r}\left(E- L\Omega\right)\partial_{r}Ldr,
\end{equation} 
where $r_{in}$ is the inner edge of the disk. The radial temperature distribution of the disk can be derived from the Stefan-Boltzmann's law

\begin{equation} \label{eq:BB_T}
F(r) = \sigma_{SB}T^4(r),
\end{equation}
where $\sigma_{SB}$  is the Stefan-Boltzmann constant. 

The observed disk luminosity has a redshifted black body spectrum:

\begin{eqnarray}
L_{\nu} = 4\pi D^2I(\nu) = \frac{8\pi h \cos{i}}{c^2}\int_{r_{in}}^{r_f}\int_{0}^{2\pi} \frac{\nu^3\sqrt{-g}drd\varphi}{exp\left(\frac{h\nu_e}{k_BT}\right) - 1},
\end{eqnarray}
where $i$ is the disk inclination angle, $D$ is the distance between the observer and the disk, $h$ is the Planck constant, $\nu_e$ is the emitted frequency , $I(\nu)$ is the Planck distribution, and $k_B$ is the Boltzmann constant.  

The observed photons are redshifted and their frequency is related to the emitted ones in the following way $\nu_e = (1+z) \nu$. The  redshift factor  $(1+z) $ has the form: 

\begin{equation}
(1+z) = \frac{1 + \Omega r \sin{\varphi} \sin{i}}{\sqrt{-g_{tt} - 2\Omega g_{t\varphi} - \Omega^2 g_{\varphi\varphi}}},
\end{equation}
where the light bending effect is neglected

Another important characteristic of the accretion disk is it's efficiency $\xi$. $\xi$ is the efficiency with which the central body converts the mass of the disk's matter into radiation. The efficiency is  measured at infinity and it is defined as the ratio of the rate of the energy of the photons emitted from the disk surface and the rate with which the mass-energy is transported to the central body. If all photons reach infinity an estimate of the efficiency is given by the specific energy measured at the ISCO:

\begin{equation}
\xi = 1 - E_{ISCO}.
\end{equation} 

The presented equations, like in the previous section, are valid both for GR and for $\mathcal{R}^2$ gravity. 

\section{Numerical results}

In this paper we are using $f(\mathcal{R})$ gravity with Lagrangian $f(\mathcal{R}) = \mathcal{R} + a\mathcal{R}^2$. This is the so-called $\mathcal{R}^2$ gravity. The equations describing slowly rotating stable stationary neutron star models  as well as rapidly rotating ones in  $\mathcal{R}^2$ gravity can be found in ~\cite{Yazadjiev2014,Staykov2014,Yazadjiev2015}. The numerical computations are made with modification \cite{Doneva2013,Yazadjiev2015} of the RNS code \cite{Stergioulas95}. The results for the presented static models are double checked with a modification of the slow rotation code used in \cite{Staykov2015a}.

In this section we concentrate either on models with equal masses and angular velocities or rotating with Kepler frequencies and have equal masses. The differences between the electromagnetic properties of the accretion disks around such stars in GR and in $\mathcal{R}^2$ gravity are investigated in detail. For the $f(\mathcal{R})$  theory we are using $a = 10^4$ which is close to the maximal allowed by observations dimensionless  value\footnote{The maximal value allowed by the observations is of order $a \sim 10^5$  or in dimensional units $a \lesssim 5\times 10^{11} \rm m^2$ \cite{Naf2010, Yazadjiev2014}.} of the free parameter in the theory The relation between the dimensional and the dimensionless values of the parameter involves one half of the gravitational radius of the Sun, $R_0 = 1.4766 \rm km$, namely $a \rightarrow a/R^2_0$.  We investigate two hadronic EOS, namely APR4 and SLy4, both of which give maximal mass higher than two solar masses in the static case, and a quark EOS -- SQS B60 with maximal mass slightly below $2M_{\odot}$ but it is a good representative nevertheless. APR4 and SLy4 are tabulated and SQS B60 has analytical form 

\begin{equation}
p = b(\rho - \rho_0),
\end{equation}
where the parameters $b$ and $\rho_0$ we take from \cite{Gondek-Rosinska2008}.

To calculate the disk parameters we are using  $\dot{M}_0 = 10^{-12} M_{\odot}  \rm year^{-1} $ \cite{Shakura1973} for the mass accretion rate. The radius of the disk is assumed to be $10^3 M$, where $M$ is the mass of the star in geometrical units. We assume the disk to be face on. The inner edge of the disk is on the ISCO, or if the ISCO is in the interior of the star the disk spreads to the  surface of the star.

\begin{table}[h]
\begin{tabular}{cccccccccc}
\hline
EOS & $\rho_{\rm c}\times 10^{15} [\rm g/cm^3]$ & $M/M_{\odot}$ & $R_{\rm e} [\rm km]$ & $\Omega [\rm s^{-1}]$ & $f = \frac{\Omega}{2\pi}$ [ Hz]& $R_{\rm ISCO} [\rm km]$\\
\hline
\multicolumn{2}{c}{GR} \\
\cline{1-2}
APR4     & 1.32    & 1.8    & 11.14    &  0 & 0& 15.96 \\
SLy4          & 1.43    & 1.8     & 11.27     &  0 & 0 & 15.95  \\
SQS B60    & 1.09   & 1.8    & 11.17    &  0 & 0 & 15.96  \\
\multicolumn{2}{c}{$f(\mathcal{R})$} \\
\cline{1-2}
APR4     & 1.21   & 1.8     &  11.79    &  0 & 0 & 14.61 \\
SLy4      & 1.26   & 1.8     &   12.01    & 0 & 0 & 14.53 \\
SQS B60    & 0.99   & 1.8     &  11.63    &  0 & 0 & 14.49 \\
\hline
\end{tabular}
\caption{Parameters of the examined static models with mass $M = 1.8M_{\odot}$ in GR and in $\mathcal{R}^2$ gravity.}
\label{Tbl:stat}
\end{table}

\begin{table}[h]
\begin{tabular}{cccccccccc}
\hline
EOS & $\rho_{\rm c}\times 10^{15} [\rm g/cm^3]$ & $M/M_{\odot}$ & $R_{\rm e} [\rm km]$ & $\Omega [\rm s^{-1}]$ & $f = \frac{\Omega}{2\pi} [\rm Hz]$ & $R_{\rm ISCO} [\rm km]$\\
\hline
\multicolumn{2}{c}{GR} \\
\cline{1-2}
APR4     & 1.225    & 1.8    & 11.77    &  5000 & 796 & 13.60 \\
SLy4          & 1.290    & 1.8     & 12.03     &  5000 & 796 & 13.65  \\
SQS B60    & 0.930   & 1.8    & 11.96    &  5000 & 796 & 13.81  \\
\multicolumn{2}{c}{$f(\mathcal{R})$} \\
\cline{1-2}
APR4     & 1.130   & 1.8     &  12.65    &  5000 & 796 & 12.65 \\
SLy4      & 1.150   & 1.8     &   13.03    & 5000 & 796 & 13.03 \\
SQS B60    & 0.855   & 1.8     &  12.47    &  5000 & 796 & 12.96 \\
\hline
\end{tabular}
\caption{Parameters of the examined models rotating with angular velocity $\Omega = 5000 \rm s^{-1}$ and mass $M = 1.8M_{\odot}$ in GR and in $\mathcal{R}^2$ gravity.}
\label{Tbl:rap}
\end{table}

\begin{table}[h]
\begin{tabular}{cccccccccc}
\hline
EOS & $\rho_{\rm c}\times 10^{15} [\rm g/cm^3]$ & $M/M_{\odot}$ & $R_{\rm e} [\rm km]$ & $\Omega [\rm s^{-1}]$ & $f = \frac{\Omega}{2\pi} [\rm Hz]$ & $R_{\rm ISCO} [\rm km]$\\
\hline
\multicolumn{2}{c}{GR} \\
\cline{1-2}
APR4     & 1.03    & 1.8    & 15.68    &  7878 & 1254  & 15.68 \\
SLy4          & 1.02    & 1.8     & 16.21     &  7504 & 1098 & 16.21  \\
SQS B60    & 0.61  & 2.0    & 17.68    &  7798 & 1241 & 19.50  \\
\multicolumn{2}{c}{$f(\mathcal{R})$} \\
\cline{1-2}
APR4     & 0.97   & 1.8     &  16.83    &  7291& 1160  & 16.83 \\
SLy4      & 0.96   & 1.8     &   17.38    & 6975 & 1110 & 17.38 \\
SQS B60    & 0.575   & 2.0     &  19.24    &  7117 & 1132 & 21.26 \\
\hline
\end{tabular}
\caption{Parameters of the examined modes rotating with Keplerian velocity and with mass $M = 1.8M_{\odot}$ for the neutron star models and  $M = 2M_{\odot}$ for the strange star models in GR and in $\mathcal{R}^2$ gravity.}
\label{Tbl:kep}
\end{table}

Our goal in this study is to compare $\mathcal{R}^2$ gravity and General Relativistic models in a way that will allow us to distinguish both theories by the differences in the emitted electromagnetic flux and the observed luminosity. Therefore,  we concentrate mainly on comparing models with  equal masses and equal angular velocities. 
In Tables \ref{Tbl:stat}, \ref{Tbl:rap}, and \ref{Tbl:kep}  the main parameters of the examined neutron and strange star models are presented. With $\rho_{\rm c}$  the central energy density of the models is marked and it is  in units of $\rm g/cm^3$. $M/M_{\odot}$ is the mass of the model, normalised to the solar mass. $R_{\rm e}$ is the equatorial radius of the modes in $\rm km$, $\Omega$ is the angular velocity in $\rm s^{-1}$, and $R_{\rm ISCO}$ is the radius of the ISCO (if the ISCO is in the interior of the star, the radius of the star is given instead) in $\rm km$. In Table \ref{Tbl:stat}  static models with mass $M = 1.8M_{\odot}$ are presented, in Table \ref{Tbl:rap} -- models with angular velocity $\Omega = 5000 \rm s^{-1}$ and mass $M = 1.8M_{\odot}$, and in Table \ref{Tbl:kep} -- models rotating with Keplerian velocity with mass $M = 1.8M_{\odot}$ for the neutron stars and  $M = 2M_{\odot}$ for the strange stars. 

\subsection{Static models}

We start our investigation of the electromagnetic properties of an accretion disk in $\mathcal{R}^2$ gravity from the case of static stars. In general, compact objects with accretion disk should rotate  due to the conservation of angular momentum and the transfer of angular momentum from the disk to the central body during accretion. However, the static models are good first approximation of the effect  that modified gravity has on the electromagnetic properties of the disk, especially if we are interested in slowly rotating stars. 

\begin{figure}[]
	\centering
	\includegraphics[width=0.45\textwidth]{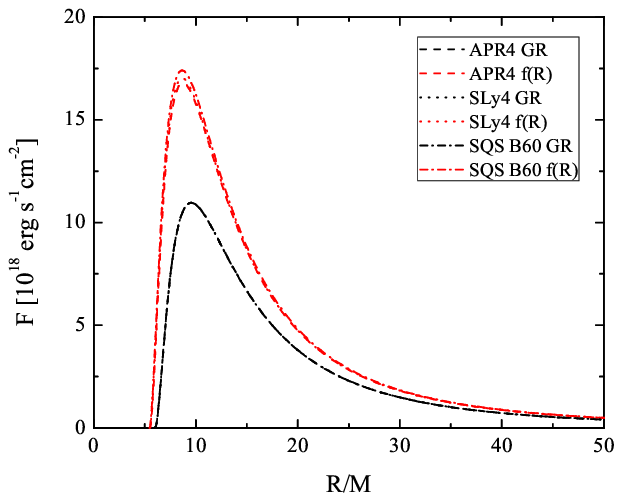}
	\includegraphics[width=0.45\textwidth]{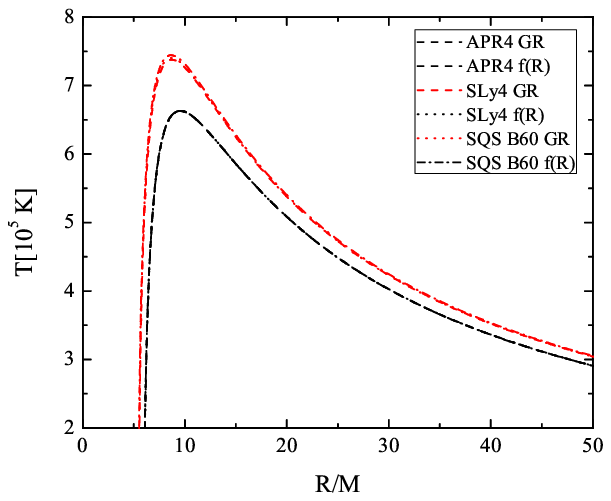}
	\caption{Electromagnetic flux  (left panel) and temperature  (right panel) as functions of the normalised to the mass of the star circumferential radius, $R/M$,  for static models with mass  $M = 1.8M_{\odot}$. The presented results are for two hadronic and one quark EOS for GR and $\mathcal{R}^2$ gravity with $a = 10^4$.}
	\label{Fig:stat_FT}
\end{figure}

In Fig. \ref{Fig:stat_FT}  the electromagnetic flux emitted from the disk and  the  effective black body temperature as functions of the reduced circumferential radius, $R/M$,  are plotted (the electromagnetic flux and the effective temperature are related by  eq. (\ref{eq:BB_T})).  In the left panel  the electromagnetic flux is plotted and in the right one -- the temperature.  The presented results are for static models  with mass $1.8M_{\odot}$ in both theories. For the $R^2$ gravity case we are using  $a = 10^4$ which is close to the maximal allowed by observations value of the parameter $a$. There are two reasons to do this. First, the deviations from the GR case are close to the maximal for this value \cite{Yazadjiev2014,Staykov2014,Yazadjiev2015}, hence it is a good starting point for investigation. The second reason is the relation between the parameter $a$ and the range around the star in which the effect of the $\mathcal{R}^2$ gravity is most pronounced. With the decrease of the value of $a$  that range rapidly decrease, so we choose a case in which the effect on the accretion disk is not negligible. 

\begin{figure}[]
	\centering
	\includegraphics[width=0.45\textwidth]{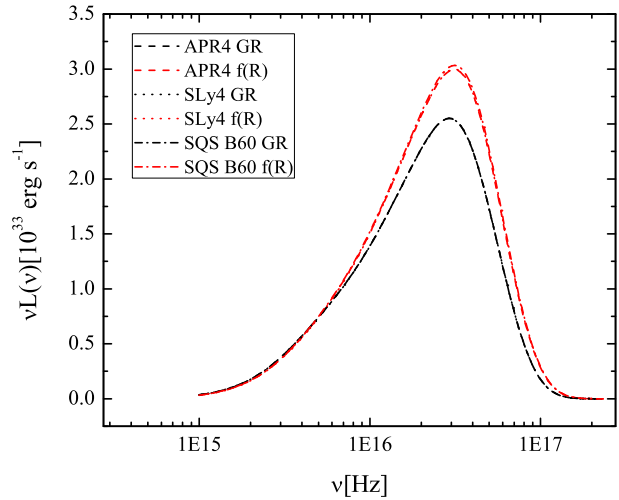}
	\caption{Luminosity as function of the observed frequency for models with mass  $M = 1.8M_{\odot}$. The presented results are for two hadronic and one quark EOS in GR and $\mathcal{R}^2$ gravity with $a = 10^4$.}
	\label{Fig:stat_L}
\end{figure}

The maximal electromagnetic flux increases with around 50\% in $\mathcal{R}^2$ gravity compared to the GR case, and the maximal temperature -- with 12\%. As expected, the effect of the modification of GR is maximal close to the  inner edge of the disk. The difference between both theories rapidly decrease with  the increase of the reduced circumferential radius and the $\mathcal{R}^2$ results converge to the GR ones away from the central body.  
In Fig. \ref{Fig:stat_L} the luminosity of the disk as a function of the reduced circumferential radius is plotted. In this case the major difference appears  for the maximal value of the luminosity -- for $\mathcal{R}^2$ gravity it increases with 20\%. The maximum of the luminosity is in the same frequency band as in GR.

\subsection{Models with equal angular velocities}

\begin{figure}[]
	\centering
	\includegraphics[width=0.45\textwidth]{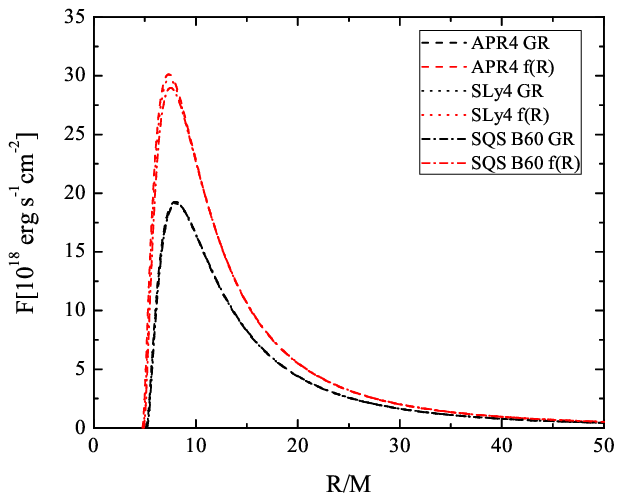}
	\includegraphics[width=0.45\textwidth]{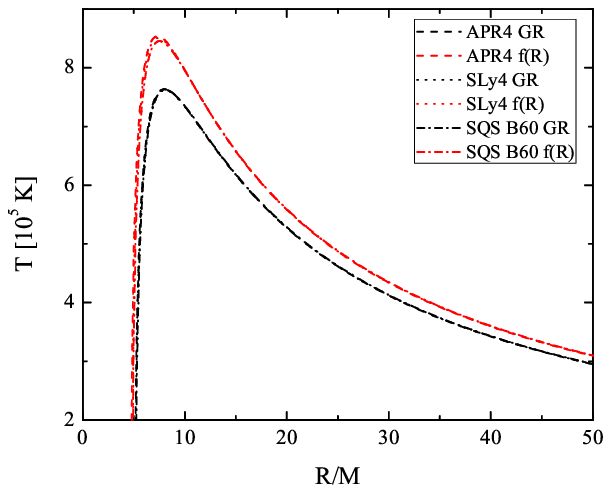}
	\caption{Electromagnetic flux  (left panel) and temperature  (right panel) as functions of the normalised to the mass of the star circumferential radius, $R/M$,  for models rotating with $\Omega = 5000 s^{-1}$ and mass  $M = 1.8M_{\odot}$. The presented results are for two hadronic and one quark EOS for GR and $\mathcal{R}^2$ gravity with $a = 10^4$.}
	\label{Fig:omega_FT}
\end{figure}

\begin{figure}[]
	\centering
	\includegraphics[width=0.45\textwidth]{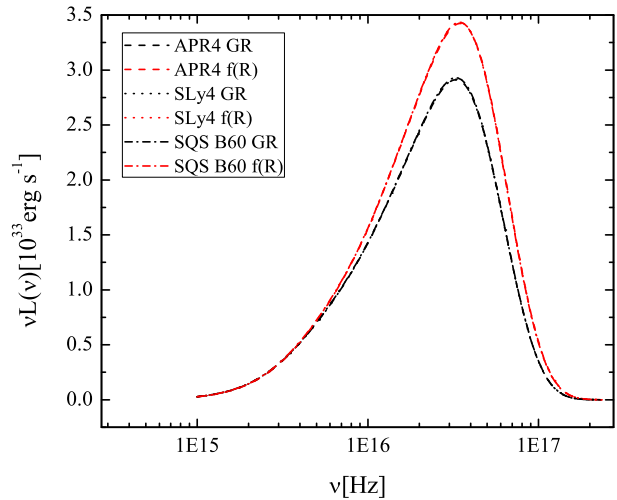}
	\caption{Luminosity as function of the observed frequency for models rotating with $\Omega = 5000 s^{-1}$ and mass  $M = 1.8M_{\odot}$. The presented results are for two hadronic and one quark EOS for GR and $\mathcal{R}^2$ gravity with $a = 10^4$.}
	\label{Fig:omega_L}
\end{figure}

The next step is to investigate rapidly rotating models. This is very interesting  and relevant case for investigation due to the fact that accretion process speeds up the central object. The fastest known rotating  compact objects are with accretion disks.  In Fig. \ref{Fig:omega_FT} and Fig. \ref{Fig:omega_L}   the results for models with mass $M = 1.8M_{\odot}$, rotating with angular velocity $\Omega = 5000 \rm s^{-1}$, which is equivalent to frequency $f = 796 \rm Hz$  are plotted. The chosen frequency is not much higher than the most rapidly rotating pulsar observed so far,  rotating with $f = 716 \rm Hz$ \cite{Hessels2006}. The values of all quantities increase with respect to the static case but the deviation of the $\mathcal{R}^2$ gravity results from the GR ones is of the same order as the static case. The flux increases with around 50 \%, the temperature with around 12 \% and the luminosity with around 20 \%.

\subsection{Models rotating with Keplerian velocity}

We continue our study with investigation of the accretion disk's electromagnetic  properties when the central body is rotating with Keplerian velocity. The examined neutron star models are with mass $M = 1.8M_{\odot}$ and the strange star models are with $M = 2M_{\odot}$. 
In this case the results are quite dispersed, compared with the previous ones. Generally the results for neutron stars are close to each other and the deviations from GR are very similar for both EOS. In $\mathcal{R}^2$ gravity the flux increases in average with around 8\%, the temperature with 2\% and the luminosity with around 10\%. The strange stars' results are more interesting. The elctromagnetic flux in this case is about 50\% lower than the neutron stars one even trough the examined models are heavier. The flux in $\mathcal{R}^2 $ gravity decreases with respect to the GR one with about 7\%, and the temperature decreases with about 2\%. The observed luminosity, however, increases with about 6\% with respect to the GR one. 

The presented in this section results deserve some additional comments. In the previous two subsections we have  found that the differences between the GR and the $\mathcal{R}^2$ gravity results are qualitatively similar for the static case and for models rotating with equal angular velocities. In this case, however, this is changed. An explanation for this behaviour we find in Table \ref{Tbl:kep}. The examined models are rotating with Kepler frequencies and they have equal masses but the rotating rate  is  higher in the GR case than the $f(\mathcal{R})$ case which compensates to a certain extend the large deviations observed in the previous sections.

\begin{figure}[]
	\centering
	\includegraphics[width=0.45\textwidth]{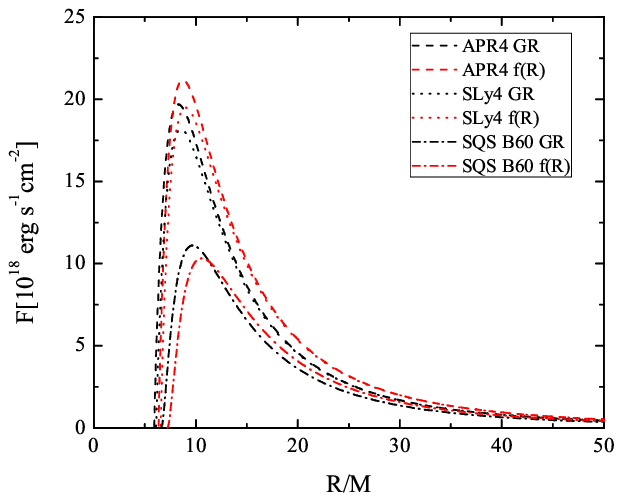}
	\includegraphics[width=0.45\textwidth]{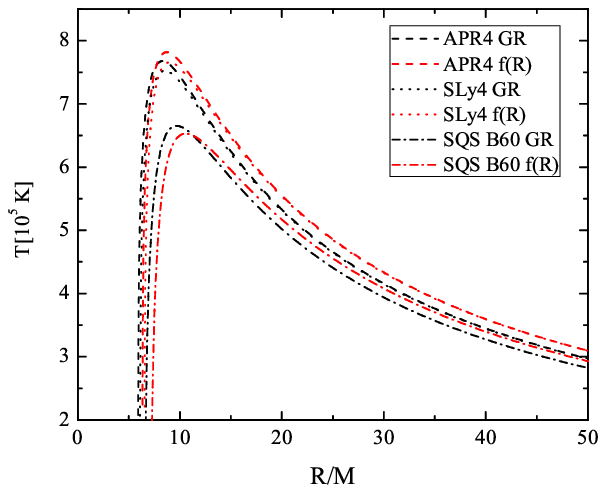}
	\caption{Electromagnetic flux  (left panel) and temperature  (right panel) as functions of the normalised to the mass of the star circumferential radius, $R/M$,  for models rotating with Keplerian velocity. The neutron stars models are with mass $M = 1.8M_{\odot}$ and the strange star models are with $M = 2M_{\odot}$. The presented results are for GR and $R^2$ gravity. }
	\label{Fig:kep_FT}
\end{figure}

\begin{figure}[]
	\centering
	\includegraphics[width=0.45\textwidth]{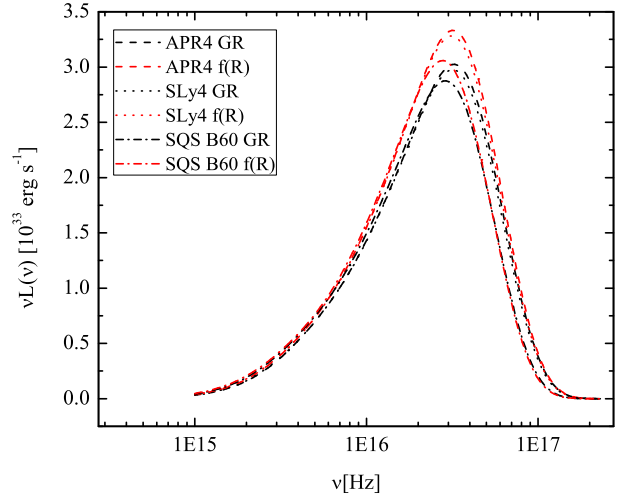}
	\caption{Luminosity as a function of the observed frequency for models rotating with Keplerian velocity.  The neutron stars models are with mass $M = 1.8M_{\odot}$ and the strange star models are with $M = 2M_{\odot}$. The presented results are for GR and $R^2$ gravity.}
	\label{Fig:kep_L}
\end{figure}

\section{Conclusions}

In the present paper we investigated the electromagnetic properties of  thin accretion disks around neutron and strange stars in GR and in $\mathcal{R}^2$ gravity with $a = 10^4$ (in our dimensional units) and commented on the differences between the results for  both theories. Static and rapidly rotating models with equal masses and equal angular velocities in both theories are investigated as well as models rotating with Keplerian velocity and equal masses in both theories. The electromagnetic flux, the temperature distribution and the luminosity are derived and compared. 

Testing alternative theories is not an easy task due to the uncertainty in our knowlge about the equation of state. In this study we demonstrated that using the accretion disk's electromagnetic properties may turn out a good way of testing the theory of gravity. We investigated the electromagnetic properties of thin accretion disks around neutron and strange stars in GR and in $\mathcal{R}^2$ gravity. The examined models are with equal masses in both theories and rotate with equal angular velocities. Our results show that the difference in the disk parameters between both theories is of the same order both for the static and rapidly rotating case considered in the paper. The increase of the electromagnetic flux is around 50\% and the increase of the luminosity of the disk is about 20 \% in the examined $f(\mathcal{R})$ theory compared to the GR. According to our results the obtained electromagnetic properties of the accretion discs around stars with equal masses and angular velocities in GR and in $\mathcal{R}^2$ gravity do not depend strongly on the EOS for the examined set of EOSs (two hadronic and one quark ones). That is why the magnetic flux and the luminosity of the disk may turn out to be excellent candidates for testing alternative theories of gravity. It is expected that the disk parameters in the same theory will be close to each other because of the similarity in the external geometry for the investigated models. This is because we used favoured by the observations modern realistic EOS.    

We studied models with  equal masses rotating with Keplertian velocity too.  The results in that case are more dispersed and the differences in between the theories decrease compared to the other considered case. This is due to the fact that examined models does not have fixed angular frequency as before, but instead the GR models are rotating with higher angular velocities than the corresponding  $\mathcal{R}^2$ gravity ones.  However, these models are not so interesting  because their angular velocity is much higher than the ones measured in the majority of the observed rotating neutron stars.

This study should be extended in several different directions. The electromagnetic properties of the accretion disks should be studied in other alternative theories. Additional EOS (softer and stiffer ones) can be added but a rough estimate on the expected results can be made by taking into account the known results in GR and the results in the present paper. The thin accretion disks are good approximation, but other more realistic disk models should be studied too.

\section*{Acknowledgements}

The authors would like to thank G. Pappas for the the discussions and the useful suggestions.
DD would like to thank the European Social Fund and the Ministry Of Science, Research and the Arts Baden-Württemberg for the support. KS and  SY would like to thank the Research Group Linkage Programme of the Alexander von Humboldt Foundation for the support. The support by the Bulgarian NSF Grant DFNI T02/6,  Sofia
University Research Fund under Grant 193/2016, and "New-CompStar" COST Action MP1304 is gratefully acknowledged.


\bibliography{references}

\end{document}